\documentclass[letterpaper]{jpconf}
\usepackage{graphicx}
\begin{document}
\title{Implications of final state interactions in $B$-decays}

\author{Alexey A Petrov}

\address{Department of Physics and Astronomy, 
        Wayne State University, Detroit, MI 48201}

\ead{apetrov@wayne.edu}

\begin{abstract}
I give a brief review of final state interactions (FSI) in heavy meson decays, paying 
particular attention to $B$-meson physics. Available theoretical methods for dealing 
with the effects of FSI are discussed.
\end{abstract}

%%%%%%%%%%%%%%%%%%%%%%%%%%%%%%%%%%%%%%%%%
\section{Introduction}

Strong interaction phases play an important role in the decays of heavy mesons. 
They produce visible effects in many nonleptonic decays and could be important 
for the proper interpretation of effects of underlying fundamental physics.
For example, strong phases between isospin amplitudes, $\delta_{1/2}$ and $\delta_{3/2}$,
\begin{eqnarray}
{\cal A}(\bar B^0 \to D^+ \pi^-)&=&\sqrt{\frac{1}{3}} |A_{3/2}| 
e^{i \delta_{3/2}} + \sqrt{\frac{2}{3}} |A_{1/2}| e^{i \delta_{1/2}} ,
\nonumber \\
{\cal A}(\bar B^0 \to D^0 \pi^0)&=&\sqrt{\frac{2}{3}} |A_{3/2}| 
e^{i \delta_{3/2}} - \sqrt{\frac{1}{3}} |A_{1/2}| e^{i \delta_{1/2}} , 
\\
{\cal A}(B^- \to D^0 \pi^-)&=&\sqrt{3} |A_{3/2}| e^{i \delta_{3/2}} ,
\nonumber
\end{eqnarray}
affect branching ratios of individual decays, as well as ratios of 
rates of isospin-related transitions. More importantly, they complicate 
interpretations of CP-violating phases from $\Delta b = 1$ transitions 
observed in the so-called direct CP-violating asymmetries. Provided
that B-decay amplitude depends on at least two amplitudes with different
weak and strong phases (for example, tree $A_1={\cal T}$ and penguin $A_2={\cal P}$
amplitudes), 
\begin{eqnarray}
{\cal A} (B \to f) &=& A_1 e^{i \phi_1} e^{i\delta_1} + 
A_2 e^{i \phi_2} e^{i\delta_2} \nonumber \\
{\cal A} (\bar B \to \bar f) &=& A_1 e^{-i \phi_1} e^{i\delta_1} + 
A_2 e^{-i \phi_2} e^{i\delta_2},
\end{eqnarray}
a CP-violating asymmetry can be formed,
\begin{equation}
A_{CP} = \frac{\Gamma(B \to f)- \Gamma(\bar B \to \bar f)}
{\Gamma(B \to f)+ \Gamma(\bar B \to \bar f)} = 
\frac{2 A_1 A_2 \sin(\phi_1-\phi_2)\sin(\delta_1-\delta_2)}
{A_1^2 + A_2^2+2 A_1 A_2 \cos(\phi_1-\phi_2)\cos(\delta_1-\delta_2)},
\end{equation}
which clearly depends on both, CP-conserving $\Delta\phi = \phi_1-\phi_2$ and 
CP-violating phase $\Delta\delta=\delta_1-\delta_2$ differences. CP-conserving 
phase difference is associated with strong interactions. There are, of course, 
many more examples. It is threfore important to have a way of computing those
phases, which in general would depend on the meson system under consideration.

The difference of the physical picture at the energy scales
relevant to $K$, $D$ and $B$ decays calls
for a specific descriptions for each class of decays.
For instance, the relevant energy scale in $K$ decays
is $m_K \ll 1$~GeV. With such a low energy release only
a few final state channels are available. This significantly
simplifies the theoretical understanding of FSI in kaon decays.
In addition, chiral symmetry can also be employed to assist
the theoretical description of FSI in $K$ decays.
In $D$ decays, the relevant scale is $m_D \sim 1$~GeV. This
region is populated by the light quark resonances, so one might expect
their significant influence on the decay rates
and $CP$-violating asymmetries. No model-independent
description of FSI is available, but it is hinted at experimentally
that the number of available channels is still limited, allowing for a
modeling of the relevant QCD dynamics.
Finally, in $B$ decays, where the relevant energy scale
$m_B \gg 1$~GeV is well above the resonance region, the heavy quark
limit might prove useful.

Final state interactions in $A \to f$ arise as a consequence of 
the unitarity of the ${\cal S}$-matrix,
${\cal S}^\dagger {\cal S} = 1$, and involve the rescattering of
physical particles in the final state.
The ${\cal T}$-matrix, ${\cal T} =  i \left (1-{\cal S} \right)$,
obeys the optical theorem:
\begin{equation}
{\cal D}isc~{\cal T}_{A \rightarrow f} \equiv {1 \over 2i}
\left[ \langle f | {\cal T} | A \rangle -
\langle f | {\cal T}^\dagger | A \rangle \right]
= {1 \over 2} \sum_{i} \langle f | {\cal T}^\dagger | i \rangle
\langle i | {\cal T} | A \rangle \ \ ,
\label{unit}
\end{equation}
where ${\cal D}isc$ denotes the discontinuity across physical cut.
Using $CPT$ in the form
$
\langle \bar f | {\cal T} | \bar A \rangle^* =
\langle \bar A | {\cal T^\dagger} | \bar f \rangle =
\langle f | {\cal T^\dagger} | A \rangle,
$
this can be transformed into 
\begin{equation}
\label{opt}
\langle \bar f | {\cal T} | \bar A \rangle^* =
\sum_{i}
\langle f | {\cal S}^\dagger | i \rangle
\langle i | {\cal T} | A \rangle \;\;.
\end{equation}
Here, the states $| i \rangle$ represent all possible final states
(including $| f \rangle$)
which can be reached from the state  $| A \rangle$
by the weak transition matrix ${\cal T}$. The
right hand side of  Eq.~(\ref{opt})
can then be viewed as a weak decay of $| A \rangle$
into $| i \rangle$ followed by a strong rescattering
of $| i \rangle$ into $| f\rangle$. Thus, we
identify $\langle f | {\cal S}^\dagger | i \rangle$
as a FSI rescattering of particles.
Notice that if $| i \rangle$ is an eigenstate of ${\cal S}$
with a phase $e^{2i\delta}$, we have
\begin{equation}
\label{triv}
\langle \bar i| {\cal T} | \bar A \rangle^* =
e^{-2i\delta_i}\langle i | {\cal T} | A \rangle \;\;.
\end{equation}
which implies equal rates for the charge conjugated
decays\footnote{This fact will be important in the studies of
$CP$-violating asymmetries as no $CP$ asymmetry is generated in 
this case.}. Also
\begin{equation}
\langle \bar i | {\cal T} | \bar B \rangle = e^{i\delta} T_i
\langle i | {\cal T} | A \rangle = e^{i\delta} T_i^*
\label{watson}
\end{equation}
The matrix elements $T_i$ are assumed to be
the ``bare'' decay amplitudes and have
no rescattering phases. This implies that these transition
matrix elements between charge conjugated states
are just the complex conjugated ones of each other.
Eq.~(\ref{watson}) is known as Watson's theorem
\cite{watson52}. Note that the problem of unambiguous
separation of ``true bare'' amplitudes from the ``true FSI'' 
ones (known as Omn\'es problem) was solved only for a
limited number of cases. 

While the above discussion gives the most general way of determining 
strong phases, especially if an S-matrix is easily diagonalized,
it might not be the most practical way of dealing with strong 
phases in $B$-decays due to the large number of available channels. 

%%%%%%%%%%%%%%%%%%%%%%%%%%%%%%%%%%%%%%%%%
\section{Decays of heavy flavors}

Theoretical analysis of decays of heavy-flavored mesons, in particular
$B$-mesons, simplifies in the limit $m_b \to \infty$. In this limit,
$q\bar q$-pair produced in the weak decay of a $b$-quark, emerges as a
small color dipole. This is a reasonable assumption, as the length scale 
of $q\bar q$ production is set by the inverse heavy quark mass, while soft 
QCD interactions are governed by the length scale associated with
$1/\Lambda_{QCD}$, and so their effects will be suppressed by powers of
$\Lambda_{QCD}/m_b$. Then, if $B\to M_1 M_2$ decay amplitude is dominated by 
this two-body-like configuration with small invariant mass, a factorization 
theorem can be written~\cite{Beneke:1999br,Beneke:2000ry,Bauer:2001cu,Politzer:1991au}
\begin{eqnarray}\label{FactTheor}
\langle M_1 M_2 | {\cal Q}_i | B \rangle &=&
\sum_j F_J^{B\to M_1} (m_2^2) \int_0^1 du ~T^I_{ij}(u) ~\Phi_{M_2}(u) +
\left(M_1 \to M_2\right) \nonumber \\
&+& \int_0^1 d\xi du dv ~T^{II}_i(\xi,u,v) ~\Phi_B(\xi) \Phi_{M_1}(v) \Phi_{M_2}(u).
\end{eqnarray}
All corrections to Eq.~(\ref{FactTheor}) should be suppressed by either
$\alpha_s$ or $1/m_b$. In fact, one can perform phenomenological 
analysis of $D$ and $B$-decays to show that amplitude behavior in
the large $m_Q$ limit is respected~\cite{Neubert:2001sj}. 

One can use perturbative arguments to calculate final state
phases for charmless $B$ decays using perturbative QCD~\cite{bss}.
Indeed, $b \to c \bar c s$ process, with subsequent final state
rescattering of the two charmed quarks into the final state (penguin
diagram) does the job, as for the energy release of the order
$m_b > 2m_c$ available in $b$ decay, the rescattered $c$-quarks
can go on-shell generating a perturbative CP-conserving phase and thus
${\cal A}_{CP}^{dir}$, which is usually quite small for the
experimentally feasible decays, ${\cal O}(1\%)$. One might be tempted 
to conclude that all strong phases in $B$-decays should be dominated by
perturbative phases from $T^I_{ij}$ and $T^{II}_i$ and therefore be small.
This conclusion, however, will not be correct if, for instance, real part
of the decay amplitude happened to be small. Since
\begin{equation}
\delta \sim \alpha_s \arctan\left({Im A \over Re A}\right),
\end{equation}
the smallness of $\alpha_s$ can be easily overpowered by large
$1/Re A$ factor. In addition, $1/m_b$ corrections could not be computed in
at this time. Yet, in this framework, it is there non-perturbative contributions to
strong phases are introduced. Those contributions could be quite large. 

A multitude of $B\to PP$ and $B \to PV$ transitions were evaluated in QCD
factorization (QCDF)~\cite{Beneke:2003zv}. However, some predictions appear in disagreement 
with experiemental data~\cite{Cheng:2004ru}. Among those, is the recent observation 
of $B \to \pi^0 \pi^0$ branching ratio, $Br(B\to \pi^0\pi^0) = (1.5\pm 0.5)\times 10^{-6}$, 
which appears to be quite larger than predicted in QCD factorizaton, 
$Br(B\to \pi^0\pi^0) = (0.3\pm 0.2)\times 10^{-6}$, inconsistensies in $B \to K\pi$ 
transitions which rule out small phases predicted by QCDF~\cite{Gronau:2006ha}, as well as 
others~\cite{Cheng:2004ru}.  

A way to estimate non-perturbative contributions is to model them using approach final state interactions
(FSI). In this approach, a transition $B \to M_1 M_2$ is divided into 
\begin{equation}
{\cal A}(B \to M_1 M_2) = {\cal A} (B \to M_1 M_2) + T(M_3 M_4 \to M_1 M_2 ) \otimes {\cal A}(B \to M_1 M_2),
\end{equation}
where $M_i$ are the final and intermediate meson states. An important point is then how to
model the rescattering amplitude $T(M_3 M_4 \to M_1 M_2)$. Most recent 
studies~\cite{Cheng:2004ru,Du:1998ss,Cheng:2005bg,Lu:2005mx} use a simple t-channel 
resonance exchange to model the rescattering. For example, $B \to \phi K$ transition
with experimental branching ratio of $Br(B \to \phi K)=(8.6\pm1.1)\times 10^{-6}$ can be
affected by a decay $B \to D_s^{(*)} D^{(*)}$ with a much larger branching ratios
of, say, $Br(B \to D_s^{*} D=(8.6\pm 3.4)\times 10^{-3}$ and subsequent rescattering of
$D_s^{(*)} D^{(*)}$ into the $\phi K$ final state via t-channel $D_s^{(*)}$-resonance 
exchange. In principle, however, other resonances, such as spin-2 $D_s^{**}$ and others 
should be taken into account. A rescattering amplitude takes a very simple form,
\begin{equation}
T(s,t)=s^J\frac{g_{M_1 M_3} g_{M_2 M_4}}{M_J^2-t},
\end{equation}
where $g_{M_i M_k}$ is a coupling constant (in practice, a $t$-dependent form-factor), and
$J$ is a spin of an exchanged particle. A big problem with this form of exchange amplitude 
is that it violates unitarity for $J \ge 2$ and thus is not appropriate for $m_b \to \infty$
limit calculations.

A solution to this problem is well-known and requires Reggeization of the scattering 
amplitudes~\cite{jc},
\begin{equation}
T(s,t)=\xi \beta(t) \left(\frac{s}{s_0}\right)^{\alpha(t)}
e^{i \pi \alpha(t)/2}
\end{equation}
with $\alpha(t)=\alpha_0 + \alpha^\prime t$ being a Regge trajectory.
One consequence of this is the fact that high energy FSIs are dominated by
the multiparticle intermediate states~\cite{dgps,dgp}.
\begin{eqnarray}
{\cal A}(B \to a b) = \sum_{c,d} T(cd \to ab )
\otimes {\cal A}(B \to cd) + \dots + \sum_{c,d,\dots ,z} T(cd \dots z \to ab )
\otimes {\cal A}(B \to cd \dots z)
\nonumber
\end{eqnarray}

It is known that scattering of high energy particles may be divided 
into `soft' and `hard' parts.  Soft scattering occurs primarily 
in the forward direction with limited transverse momentum 
distribution which falls exponentially with a scale of order 
$0.5$~GeV. At higher transverse momentum one encounters the 
region of hard scattering, which can be described by 
perturbative QCD. In exclusive $B$ decay 
one faces the difficulty of separating the two.
It might prove useful to employ unitarity in trying to 
describe FSI in exclusive $B$ decays. 

It is easy to investigate first the {\it elastic} channel.
The inelastic channels have to share a similar asymptotic
behavior in the heavy quark limit due to the unitarity of the 
${\cal S}$-matrix. The choice of elastic channel 
is convenient because of the optical theorem which connects the 
forward (imaginary) invariant amplitude 
${\cal M}$ to the total cross section,
\begin{equation}
{\cal I}m~{\cal M}_{f\to f} (s, ~t = 0) = 2 k
\sqrt{s} \sigma_{f \to {\rm all}} \sim s \sigma_{f \to {\rm all}} \ \ ,
\label{opti}
\end{equation}
where $s$ and $t$ are the usual Mandelstam variables. The asymptotic 
total cross sections are known experimentally to rise slowly 
with energy and can be parameterized by the form \cite{pdg},
$
\sigma (s) = X \left({s/s_0}\right)^{0.08}
+ Y \left({s/s_0}\right)^{-0.56},
$
where $s_0 = {\cal O}(1)$~GeV is a typical hadronic scale.
Considering only the imaginary part of the amplitude, and building in 
the known exponential fall-off of the elastic cross section in $t$  
($t<0$)~\cite{jc} by writing
\begin{equation}
i{\cal I}m~{\cal M}_{f\to f} (s,t) \simeq i \beta_0 \left( {s \over s_0}
\right)^{1.08} e^{bt} \ \ ,
\label{fall}
\end{equation}
one can calculate its contribution to the
unitarity relation for a final state $f = ab$ with kinematics
$p_a' +  p_b' = p_a +  p_b$ and $s = (p_a + p_b )^2$:
\begin{eqnarray}
{\cal D}isc~{\cal M}_{B \to f} &=&
{-i \over 8 \pi^2} \int {d^3p_a' \over 2E_a'}
{d^3p_b' \over 2E_b'}
\delta^{(4)} (p_B - p_a' - p_b') 
{\cal I}m~{\cal M}_{f\to f} (s,t)
{\cal M}_{B \rightarrow f} \nonumber \\
&=& - {1\over 16\pi} {i\beta_0 \over s_0 b}\left( {m_B^2 \over s_0} 
\right)^{0.08} {\cal M}_{B \rightarrow f} \ \ ,
\label{mess}
\end{eqnarray}
where $t = (p_a - p_a')^2 \simeq -s(1 - \cos\theta)/2$,
and $s = m_B^2$.

One can refine the argument further, since
the phenomenology of high energy
scattering is well accounted for by the Regge theory~\cite{jc}.
In the Regge model, scattering amplitudes are described by the 
exchanges of Regge trajectories (families of particles of differing 
spin) with the leading contribution given by the Pomeron
exchange. Calculating the Pomeron contribution to the
elastic final state rescattering in $B \to \pi \pi$
one finds \cite{dgps}
\begin{equation}
{\cal D}isc~{\cal M}_{B \to \pi\pi}|_{\rm Pomeron} = -i\epsilon
{\cal M}_{B \to \pi\pi}, ~~~~~
\epsilon \simeq 0.21 \ \ .
\label{despite}
\end{equation}
It is important that the Pomeron-exchange amplitude is seen to be almost 
purely imaginary. However, of chief significance is the
identified weak dependence of $\epsilon$ on $m_B$  -- the
$(m_B^2)^{0.08}$ factor in the numerator is attenuated by the
$\ln(m_B^2/s_0)$ dependence in the effective value of $b$.

The analysis of the elastic channel suggests that, at high energies,  
FSI phases are {\it mainly generated by inelastic effects}, which
follows from the fact that
the high energy cross section is mostly inelastic. This also
follows from the fact that the Pomeron elastic  
amplitude is almost purely imaginary.  Since the study of
elastic rescattering has yielded a ${\cal T}$-matrix element ${\cal  
T}_{ab
\to ab} = 2 i \epsilon$, i.e. ${\cal S}_{ab \to ab} = 1- 2
\epsilon$, and since the constraint of unitarity of
the ${\cal S}$-matrix
implies that the
off-diagonal elements are ${\cal O}(\sqrt{\epsilon})$,
with $\epsilon$
approximately ${\cal O}(m_B^0)$ in powers of $m_B$ and numerically
$\epsilon < 1$, then the inelastic amplitude must also be ${\cal
O}(m_B^0)$ with $\sqrt{\epsilon} > \epsilon$.
Similar conclusions follow from the consideration of 
the final state unitarity relations. This complements the old 
Bjorken picture of heavy meson decay (the dominance of the
matrix element by the formation of the 
small hadronic configuration which grows into the final state
pion ``far away'' from the point it was produced and does not 
interact with the soft gluon fields present in the decay, see 
also~\cite{DP97} for the discussion) by allowing 
for the rescattering of multiparticle states, production of
whose is favorable in the $m_b \to \infty$ limit, into the
two body final state.
Analysis of the final-state unitarity relations in their general 
form is complicated due to the many contributing intermediate 
states, but we can illustrate the systematics of inelastic 
scattering in a two-channel model. It involves a two-body final 
state $f_1$ undergoing elastic scattering and a final state $f_2$ 
which represents `everything else'. As before, the elastic amplitude 
is purely {\it imaginary}, which dictates the following 
{\it one-parameter} form for the $S$ matrix
\begin{equation}
 S = \left( \begin{array}{cc} 
\cos 2 \theta & i \sin 2 \theta \\
              i \sin 2 \theta & \cos 2 \theta
\end{array} \right)  \ ,\qquad \qquad 
 T = \left( \begin{array}{cc}
2 i \sin^2 \theta &  \sin 2 \theta \\
               \sin 2 \theta & 2 i \sin^2 \theta 
\end{array} \right)  \ \ ,
\label{matr1}
\end{equation}
where we identify $\sin^2 \theta \equiv \epsilon$. The unitarity 
relations become
\begin{eqnarray}
{\cal D}isc ~{\cal{M}}_{B \to f_1} &=& 
- i \sin^2 \theta {\cal{M}}_{B \to f_1} +
\frac{1}{2} \sin 2 \theta {\cal{M}}_{B \to f_2} \ \ ,\nonumber \\
{\cal D}isc~ {\cal {M}}_{B \to f_2} &=& \frac{1}{2} \sin 2 \theta 
{\cal{M}}_{B \to f_1} - i \sin^2 \theta {\cal{M}}_{B \to f_2} \ \ 
\label{big}
\end{eqnarray}
Denoting ${\cal{M}}_1^0$ and ${\cal{M}}_2^0$ 
to be the decay amplitudes in the limit $\theta \to 0$,
an exact solution to Eq.~(\ref{big}) is given by
\begin{equation}
{\cal{M}}_{B \to f_1} = \cos \theta {\cal{M}}_1^0 + i \sin \theta
{\cal{M}}_2^0  \ , \qquad 
{\cal{M}}_{B \to f_2} = \cos \theta {\cal{M}}_2^0 + i \sin \theta
{\cal{M}}_1^0 \ \ .
\label{soln}
\end{equation}
Thus, the phase is given by the inelastic scattering with a 
result of order 
\begin{equation}
{\cal I}m~ {\cal{M}}_{B \to f}/{\cal R}e~ {\cal{M}}_{B \to f} 
\sim 
\sqrt{ \epsilon}~ \left({\cal{M}}_2^0/{\cal{M}}_1^0\right) \ \ .
\end{equation}
Clearly, for physical $B$ decay, we no longer 
have a simple one-parameter ${\cal S}$ matrix, and,
with many channels, cancellations or enhancements are 
possible for the sum of many contributions.
However, the main feature of the above result is expected 
to remain: inelastic channels cannot vanish and provide the 
non-perturbative FSI phase. Further studies of flavor-nondiagonal transitions
showed that FSIs are suppressed by powers of $1/m_b$~\cite{Falk:1998wc,Chua:2002wk,Kaidalov:2006uf} 
and thus can be used to estimate nonperturbative strong phases in the 
heavy quark limit.

%%%%%%%%%%%%%%%%%%%%%%%%%%%%%%%%%%%%%%%%%%%%%%%
\section{Conclusions} 
I reviewed the physics of final state interactions 
in meson decays. I showed that the use of pole exchange graphs to
model FSI amplitudes leads to violation of unitarity in the large $m_b$ limit,
so consistent studies of high-energy rescattering amplitudes must 
involve Reggiazed rescattering amplitudes. Several examples were provided.

One of the main goals of physics of $CP$ violation and 
meson decay is to correctly extract the underlying parameters of 
the fundamental Lagrangian that are responsible for these
phenomena. The understanding of final state interactions 
is very important for the success of this program.

%%%%%%%%%%%%%%%%%%%%%%%%%
\section{Acknowledgments}
This work was supported in part by the U.S.\ National Science Foundation under
CAREER Award PHY--0547794, and by the U.S.\ Department of Energy 
under Contract DE-FG02-96ER41005.

%%%%%%%%%%%%%%%%%%%%%%%%%%%%%%%%%%%%%%%%%%%%%%

\end{document}